\documentclass[useAMS,usenatbib]{mn2e}
\usepackage{graphicx,natbib,amsmath}

\title[Conjugate-plane photometry]{Conjugate-plane photometry: Reducing scintillation in ground-based photometry}	

\author[J. Osborn et al.]{James Osborn$^1$, Richard W.~Wilson$^1$, V.~S. Dhillon$^2$, Remy Avila$^3$
\newauthor and Gordon D.~Love$^1$\\	 
$^1$Department of Physics, Centre for Advanced Instrumentation, University of Durham, South Road, Durham DH1 3LE\\
$^2$Department of Physics and Astronomy, University of Sheffield, Sheffield, S3 7RH\\
$^3$Centro de F\`{i}sica Aplicada y Tecnolog\`{i}a Avanzada, Universidad Nacional Aut\`{o}noma de M\`{e}xico, A.P. 1-1010, Santiago de Quer\`{e}taro,\\
 Quer\`{e}taro 76000, M\`{e}xico}

\newcommand{\comment}[1]{}
\begin{document}
\maketitle	
\begin{abstract}
High precision fast photometry from ground-based observatories is a challenge due to intensity fluctuations (scintillation) produced by the Earth's atmosphere. Here we describe a method to reduce the effects of scintillation by a combination of pupil reconjugation and calibration using a comparison star. Because scintillation is produced by high altitude turbulence, the range of angles over which the scintillation is correlated is small and therefore simple correction by a comparison star is normally impossible. We propose reconjugating the telescope pupil to a high dominant layer of turbulence, then apodizing it before calibration with a comparison star. We find by simulation that given a simple atmosphere with a single high altitude turbulent layer and a strong surface layer a reduction in the intensity variance by a factor of $\sim30$ is possible. Given a more realistic atmosphere as measured by SCIDAR at San Pedro M$\acute{\mathrm{a}}$rtir we find that on a night with a strong high altitude layer we can expect the median variance to be reduced by a factor of $\sim11$. By reducing the scintillation noise we will be able to detect much smaller changes in brightness. If we assume a 2~m telescope and an exposure time of 30 seconds a reduction in the scintillation noise from 0.78~mmag to 0.21~mmag is possible, which will enable the routine detection of, for example, the secondary transits of extrasolar planets from the ground.
\end{abstract}
\section{Introduction}

High precision fast photometry is key to several branches of research including (but not limited to) the study of extrasolar planet transits (e.g. \citeauthor{Charbonneau2000} 2000), stellar seismology \citep{Dalsgaard07} and the detection of small Kuiper belt objects (e.g. \citeauthor{Schlichting09} 2009). The difficulty with such observations is that, although the targets are often bright, the variation one wishes to detect\comment{amplitude of variability} is often very small (typically millimagnitudes or less) and hence the signal to noise ratio is not limited by the detector or sky but by intensity fluctuations (scintillation) produced by the Earth's atmosphere. For this reason fast photometers are generally put in space (e.g. CoRoT, Kepler and PLATO).

Extrasolar planetary transits can be detected from the ground. However the measurement of the secondary eclipse (i.e. where the planet goes behind the star) is a challenge. Such observations are crucial, as only the secondary eclipse can give information on the planetary atmosphere, including the temperature and albedo \citep{Knutson07}. Secondary eclipses were detected for the first time from space in 2005 using Spitzer at 3~$\mu$m \citep{Charbonneau05}. There has been a great deal of effort to detect secondary eclipses from the ground, but for years no detections were made (in large part due to scintillation noise). Finally, in 2009, the first ground-based detections were made, but these relied on near-IR measurements and had to target the most bloated, closest (to their host star) exoplanets to maximise the eclipse signal \citep{Sing09}. Since then a few other exoplanets have had secondary eclipses detected from the ground in this way. As noted by \citet{Deming09}, secondary eclipses recorded in visible light in addition to IR measurements are crucial if we are to understand the relative contribution of thermal emission and reflected light, and the planetary albedo.

\comment{
and the planet light is expected to peak at much bluer wavelengths (in the red/near-IR). 

Transits, of course, can easily be detected from the ground using even small amateur telescopes. The really tricky thing is to detect the secondary eclipses (or occultations), i.e. where the planet goes behind the host star. 

Our technique promises to:
1. Extend such studies to less extreme systems (i.e. the vast majority) which have weaker eclipse signals.
2. Most importantly, extend such studies to optical wavelengths. As noted by Deming and Seger in their recent Nature review (http://www.nature.com/nature/journal/v462/n7271/full/nature08556.html), under the "New observing windows and techniques" section, secondary eclipses recorded in visible light are crucial if we are to understand the relative contribution of thermal emission and reflected light, and the planetary albedo.
}

Time averaging the intensity will reduce the scintillation noise by an amount proportional to the square root of the exposure time \citep{Dravins1998}, but this will often result in saturating the CCD which then requires de-focusing the telescope to distribute the image of the star over more pixels. De-focusing has certain advantages, such as reducing the impact of pixel-to-pixel and intra-pixel sensitivity variations, but it also significantly increases the sky and CCD readout noise \citep{southworth2009}. In addition, de-focusing is not routinely possible on some telescopes (e.g the VLT) \comment{active optics WFS}and it can not be done with crowded fields. More importantly for fast photometry, time averaging can also only be used in circumstances where the intrinsic variability of the target has a much longer time scale than the scintillation. 
As scintillation is caused by the spatial intensity fluctuations crossing the pupil boundary, the time scale is determined by the wind speed of the turbulent layer. \citet{Dravins1997,Dravins1997b,Dravins1998} studied the temporal autocorrelation of the scintillation pattern at astronomical sites and found that the power is mainly located between 10 and 100~Hz but actually spans many orders of magnitude. 

Differential photometric measurements can be made by normalising with a nearby comparison star (e.g. \citeauthor{Henry2000} 2000). This is not to reduce the scintillation but to correct  for transparency variations in the atmosphere. However, this actually makes the scintillation noise worse as it is inherently caused by high altitude layers and therefore will have a very small angle of coherence \comment{(defined here as the isophotometric angle, analogous to the isoplanatic angle for wavefront phase) }in the optical (typically $\sim1^{\prime\prime}$). Here we propose a technique, called ``conjugate-plane photometry'' to reduce scintillation noise by increasing the \comment{isophotometric }angle of coherence up to $\sim0.5^{\circ}$, allowing the intensity variations of the target star to be corrected by a comparison star. Our technique offers a relatively simple way of routinely obtaining space-quality photometry from the ground for a fraction of the price and with much larger telescope apertures.

In section 2 we describe the scintillation reduction method. Section 3 shows the results of simulations of our correction technique. The expected performance of the system for a theoretical extrasolar planet transit, and simulation results using a real atmospheric profile are shown in section 4. Finally in section 5 we discuss the design of a prototype which will be tested at the NOT on La Palma in September 2010.

\section{Scintillation Calibration}

High altitude turbulence in the atmosphere distorts the plane wavefronts of light from a star which is effectively at infinity. As the wavefronts propagate these phase aberrations evolve into intensity variations which we view with the naked eye as twinkling. Wavefronts incident on a telescope pupil have both phase variations, caused by the integrated effect of light passing though the whole vertical depth of the atmosphere, and intensity variations, caused predominantly by the light diffracting through high altitude turbulence and interfering at the ground. Phase variations are normally considered more significant as they dramatically affect the spatial resolution of images, and this has led to the development of adaptive optics.  The intensity variations across the pupil are effectively averaged together when the light is focused and therefore have  less effect. A larger aperture implies more spatial averaging (which is why
stars twinkle less when observed through a telescope than with the naked eye).
However, these small intensity fluctuations do become significant when one is concerned with high precision photometry. 

\begin{figure}
	\centering
	 \includegraphics[width=80mm]{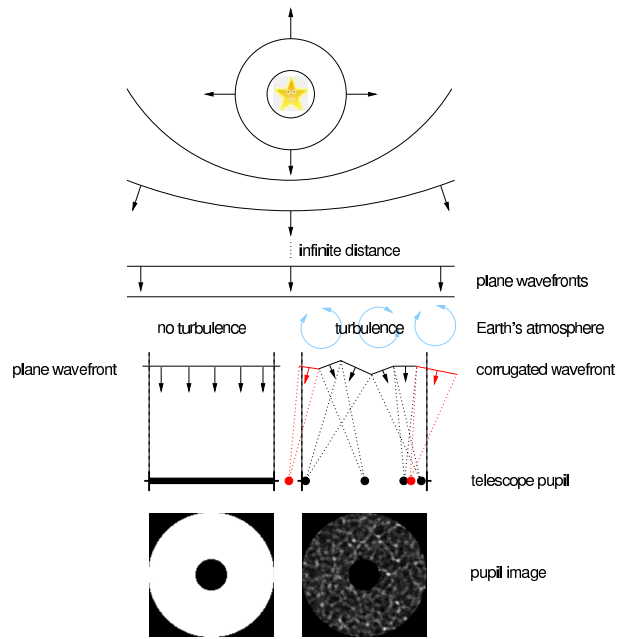} 
	 \caption{A spherical wavefront from a star will appear flat as it enters the the atmosphere. In the absence of turbulence this flat wavefront will be collected by the telescope pupil (left). In the presence of turbulence the wavefront will diffract through the refractive index variations which accompany the turbulent motion in the atmosphere. The wavefront will then interfere with itself at the ground and cause intensity fluctuations. A simplified geometrical model is shown on the right. The scintillation noise occurs when extra light is focused into the telescope pupil or when light is focused away from the pupil by the turbulent atmosphere.}
	 \label{fig:scintillation1}
\end{figure}

Consider now the effect of these intensity variations in more detail.  If we ignore diffraction, then a flat wavefront which is the same size as the telescope pupil at a given high altitude, in the absence of atmospheric turbulence, will propagate in a direction normal to the wavefront and will be collected by the telescope pupil (see figure~\ref{fig:scintillation1}). Now consider the effect of atmospheric distortion. Phase aberrations cause diffraction in different directions and hence produce scintillation. Effectively light from one part of the original wavefront is redirected to other parts of the pupil. This in itself is not a significant problem for photometry, as the integrated intensity across the pupil is the same. The problem occurs either when rays from the wavefront at high altitude propagate away from the telescope pupil, and are lost, or conversely when high altitude areas away from the telescope pupil area propagate into the telescope pupil at the ground. These effects lead to a decrease and increase in intensity, respectively, and at any one instant both of these effects will be occurring (see figure~\ref{fig:scintillation1}). The turbulence is blown across the field of view of the telescope producing an overall change in intensity as a function of time.

\begin{figure}
	\centering
	 \includegraphics[width=50mm]{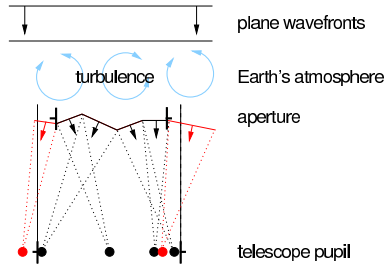} 
	 \caption{By placing an aperture at the altitude of the turbulent layer we can reduce the scintillation noise. It will now be impossible for any light from outside of the telescope pupil to be focused into the collecting area. It will also be unlikely for any parts of the wavefront to be focused off-axis by such a degree as to escape from the collecting area all together. These two situations are shown in red. These rays  - which would normally
  be the ones producing a change in the overall intergrated intensity - are blocked.}
	 \label{fig:scintillation2}
\end{figure}

As a thought experiment, to show the basic concept behind our proposal, if we could place an aperture which is smaller than the telescope pupil in the sky at the altitude of high turbulence then this change in intensity could be dramatically reduced. In this case, the rays that would have been deflected away from the area of the pupil would still be collected by the (larger) telescope pupil, and as the angle of diffraction is small no rays would be deflected into the telescope pupil because of the aperture (see figure~\ref{fig:scintillation2}). 
Increasing the size difference between the aperture in the sky and the telescope pupil would
improve the scintillation rejection, but would also lead to increased loss of signal, and clearly a balance
between the two effects would need to be found.

\begin{figure}
	\centering
	 \includegraphics[width=70mm]{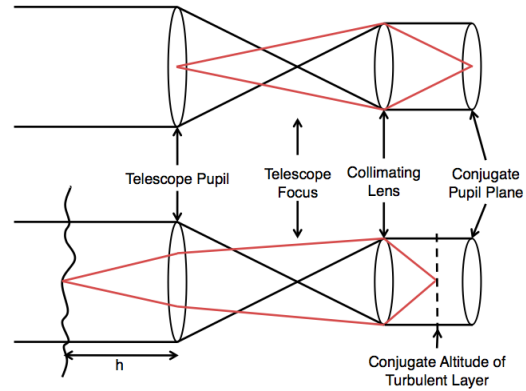} \\
	 \caption{Ray diagrams for conjugation positions. The black lines show the rays for an object at infinity. The top diagram shows the conjugate position of the telescope pupil. Every point in this plane will be an image of a point on the telescope pupil (as shown by the red lines). The lower diagram shows that by moving the observation plane towards the collimating lens then an image of the wavefront at a height $h$ above the telescope will be produced. If a camera is in a position such that it is in the image plane of the turbulent layer it is at the conjugate altitude of that layer. In practice subsidiary optics may also be used, but this diagram shows the principle.}
	 \label{array:recon}
\end{figure}

Placing an aperture at a high altitude in the sky is clearly an impractical proposal, but we can produce a similar effect using optics after the 
telescope focus. Figure~\ref{array:recon} shows how reconjugation can be produced by
 observing the beam in a different plane downstream from the telescope focus. The high altitude turbulent layer is reimaged onto an aperture which is slightly smaller than the equivalent size of the full telescope pupil. Consider again the simplified case of a single layer of turbulence at a high altitude. As already described, this produces scintillation in the entrance pupil of the telescope. If we reimage the high altitude layer at a conjugate plane then the rays will have propagated so as to ``undo'' the scintillation and we would view an approximately uniform intensity \citep{Fuchs98}. High altitude areas of the wavefront, which in the absence of turbulence would fall outside of the telescope pupil, can be diffracted by the turbulence and interfere to cause intense regions within the pupil area. This light would image in the conjugate plane outside of the aperture and can be easily rejected by the mask. High altitude areas of the wavefront which are diffracted by the turbulence and interfere to cause intense areas at the ground outside of the telescope pupil are lost and will show up as areas of decreased intensity towards the edge of the reimaged wavefront. This effect can also be rejected with a mask at the reimaged altitude which is slightly smaller than the pupil size. The remaining light within this mask will be approximately of uniform intensity and scintillation free.

\begin{figure}
	\centering
	 \includegraphics[width=80mm]{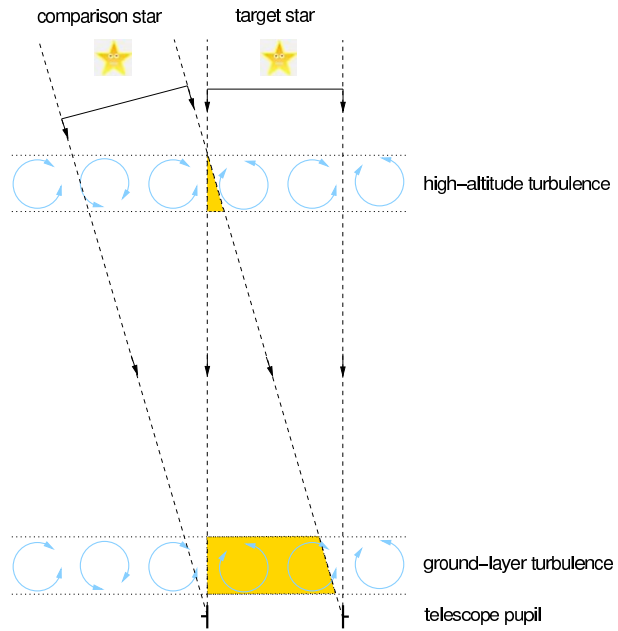} 
	 \caption{In differential photometry the intensity of the target star is calibrated by the intensity from a second comparison star. As the scintillation is caused by high altitude turbulence the two light cylinders do not sample the same turbulence and hence there will be very little correlation between the two. By conjugating the telescope to the high-altitude layer we remove the scintillation from this layer and it is replaced by scintillation from the surface turbulent layer instead. However, as the two light cylinders sample the same region of turbulence near the ground they will have very similar scintillation patterns, allowing one to be corrected by the other. The angle of separation of the two stars can be large as the surface layer is generally found to be thin.}
	 \label{fig:scintillation3}
\end{figure}

The above description has ignored two important effects, namely diffraction and turbulence from other atmospheric layers (predominantly low altitude turbulence). As well as high altitude turbulence most astronomical sites will have a strong surface layer \citep{Osborn10,Chun09} and possibly turbulence at intermediate altitudes as well. If we conjugate our system to the altitude of a high turbulent layer we will still see scintillation from other layers. We will have effectively swapped scintillation caused by high altitude 
turbulence with scintillation caused by turbulence close to the ground. 
\citet{Fuchs98} demonstrated that if a turbulent layer is below the conjugate plane (the surface layer for example) then a virtual reverse propagation occurs over a distance $z = | h - z_{0} |$, where $z_{0}$ is the conjugate altitude and $h$ is the altitude of the turbulent layer. Therefore the surface layer will now cause scintillation in the conjugate plane as it will have effectively propagated a distance $z_{0}$. However a comparison star can be used to reduce the scintillation from the surface layer as they will both sample the same turbulent area, as shown in figure~\ref{fig:scintillation3}. This layer must also be quite thin to ensure the wavefront samples the same turbulence, and studies have demonstrated that this is the case (it is often only a few 10's of meters, \citeauthor{Osborn10}, 2010, \citeauthor{Tokovinin10}, 2010, \citeauthor{Chun09}, 2009) meaning that the coherence angle is now very large (up to $0.5^{\circ}$).

\begin{figure}
	\centering
	$\begin{array}{cc}
	 \includegraphics[width=38mm]{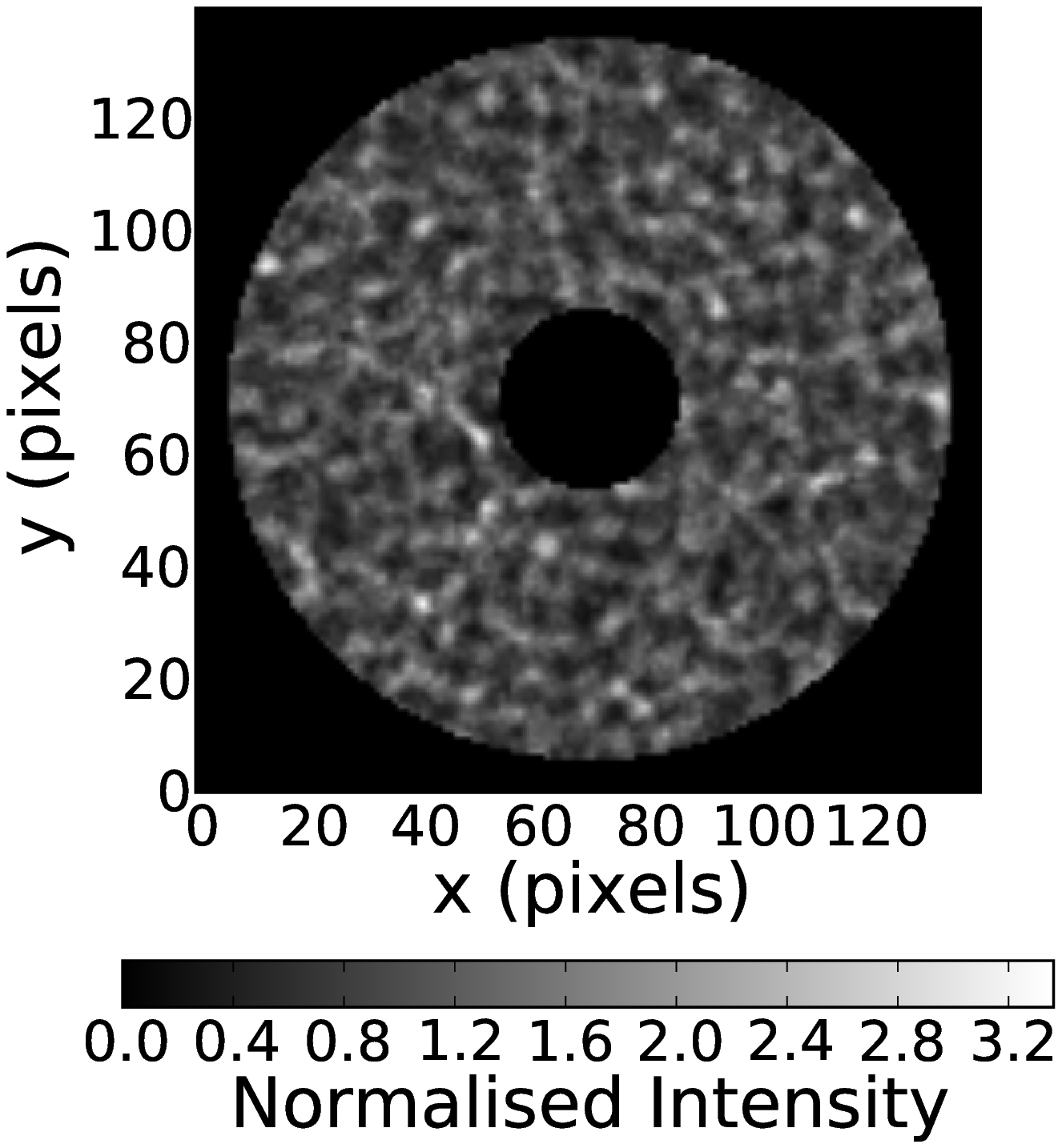} &
	\includegraphics[width=38mm]{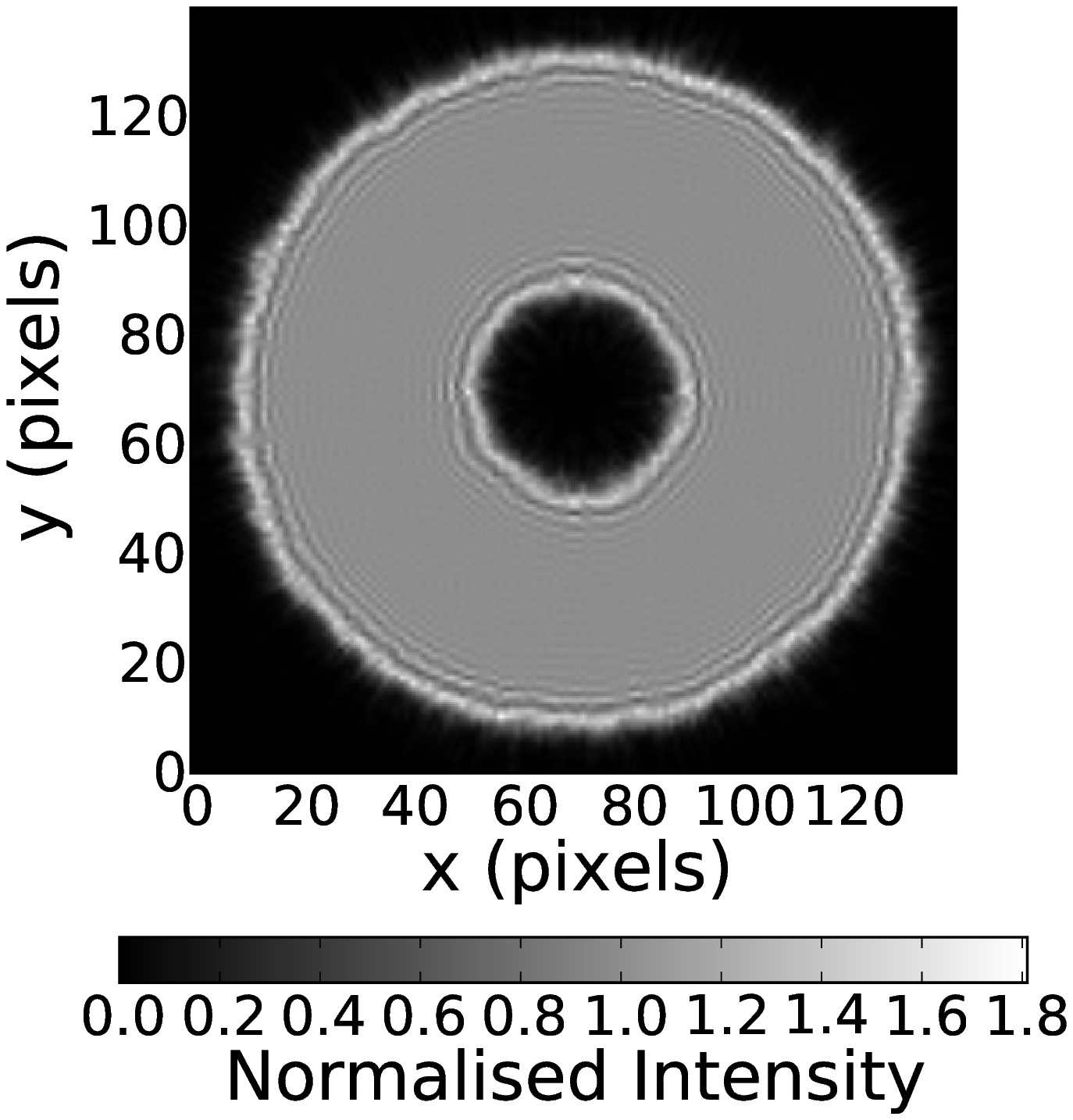} \\ 
	 \end{array}$
	 \caption{Simulated pupil intensity patterns at the telescope pupil (left) and at the conjugate altitude of the turbulent layer (right). The telescope pupil is 2.0~m in diameter and the turbulent layer has $\int C_n^2 dh = 353\times10^{-15}~ \rm{m}^{1/3}$\comment{ $r_0$ of 0.15~m} and is located at an altitude of 10~km. The intensity pattern at the conjugate altitude shows that the spatial intensity fluctuations have been removed but have been replaced by diffraction rings concentrated around the edges that also permeate throughout the pupil.}
	 \label{array:scint_pupil}
\end{figure}

\begin{figure}
	\centering
	$\begin{array}{cc}
	 \includegraphics[width=38mm]{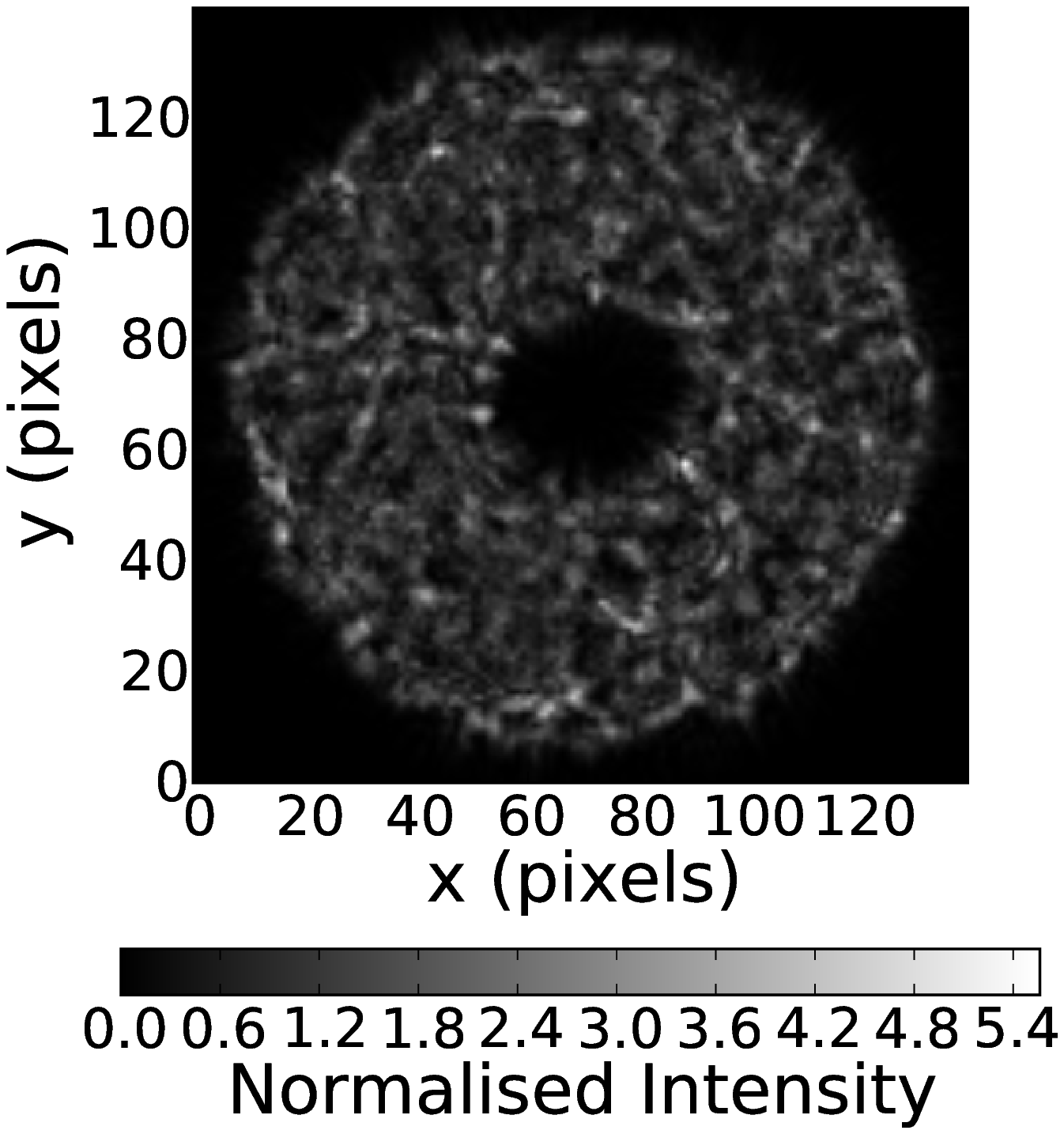} &
	 \includegraphics[width=38mm]{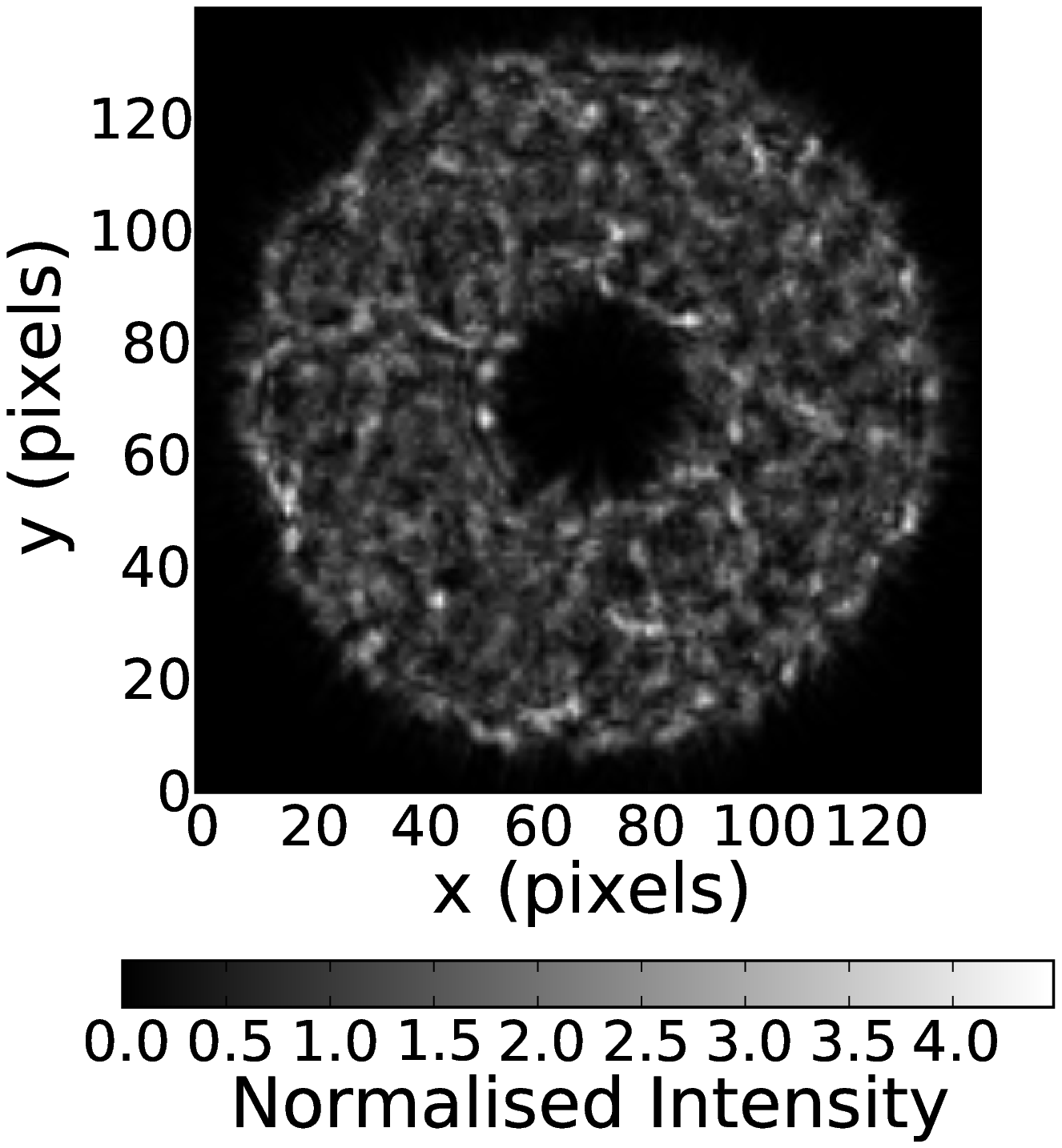}
	 \end{array}$
	 \caption{Pupil images conjugate to 10~km for two stars separated by 40$^{\prime\prime}$. The spatial intensity fluctuations are a combination of the scintillation pattern from the surface turbulent layer and the diffraction pattern of the telescope pupil (figure~\ref{array:scint_pupil}, right). The two images have very similar intensity patterns as they are both formed by the propagation of the same area of surface layer.}
	 \label{fig:rpup}
\end{figure}

Figure~\ref{array:scint_pupil} shows the effect of reconjugation of a single high altitude layer, including the effects of diffraction caused by the telescope pupil. The simulation assumed a single high altitude turbulent layer at 10~km with $\int C_n^2 dh = 353\times10^{-15}~ \rm{m}^{1/3}$, where $C_n^2$ is the refractive index structure constant and $\int C_n^2 dh$ is the integrated turbulence strength of the atmospheric layer. This corresponds to $r_{0}=0.15~\rm{m}$, where $r_{0}$ is the Fried parameter and is a measure of the integrated strength of the turbulence. It can be seen that the variations in intensity due to scintillation largely disappear in the reconjugated image of the high altitude layer - but that diffraction can clearly be seen. The diffraction rings are not completely circular as a result of the phase distortions in the wavefront at the telescope pupil. Figure~\ref{fig:rpup} shows simulated images of the reconjugated pupils at 10~km for a two-layer atmosphere (0 and 10~km) for two stars separated by 40$^{\prime\prime}$. The two images are very similar indicating that one may be used to calibrate the other. They are not identical, however, as they are being not being illuminated by a flat, uniform wavefront due to the high altitude turbulence (and not the finite thickness of the layer), and this introduces a source of error - as highlighted in the next section.


\section{Theory and Simulation Results}

Assuming a single turbulent layer at 10~km and no other turbulence the wavefunction, $\Psi$, at the telescope pupil is given by,
\begin{equation}
    \Psi(x,y) =\left[ K(z=+10~\mathrm{km}) \otimes \exp{(i\phi_{10})} \right] P(x,y),
\end{equation}
where $z$ is the propagation distance, $x$ and $y$ are spatial co-ordinates, $P(x,y)$ is the telescope pupil function, $\phi_{h}$ is the turbulent phase screen at altitude $h$~km, $\otimes$ denotes a convolution and $K$ is the Fresnel propagation kernel, given by,
\begin{equation}
K = \frac{i}{\lambda z} \exp\left(ikz\right) \exp\left(\frac{ik}{2z} \left[ \left(x-x^\prime\right)^{2} + \left(y-y^\prime\right)^{2} \right] \right),
\end{equation}
where $k$ is the wavenumber $\lambda$ is the wavelength of the light and $x^\prime$ and $y^\prime$ and spatial co-ordinates in the observation plane located at a distance $z$. Positive $z$ indicates a diverging spherical wavefront and negative $z$ is a converging spherical wavefront or a negative propagation. Therefore, the wavefunction in the conjugate plane, $\Psi^{\prime}(x^\prime,y^\prime)$, is found by a further propagation of the wavefront by a negative distance,
\begin{multline}
    \Psi^{\prime}(x^\prime,y^\prime) = \\K(z=-10~\mathrm{km}) \otimes \left[ \left[ K(z=+10~\mathrm{km}) \otimes \exp{(i\phi_{10})} \right] P(x,y) \right].
\end{multline}
In the case of an infinitely large pupil, $\Psi^{\prime}(x^\prime,y^\prime) = \Psi(x,y)$ and the pupil amplitude is flat. Therefore, by placing the aperture at the conjugate altitude of the turbulent layer we can reduce the scintillation caused by that layer. However, with a real aperture the intensity profile at the conjugate plane is not flat because the wavefront diffracts through the telescope pupil and causes diffraction rings at the edge of the pupil image which are a function of the turbulent phase screen. If we include a ground layer, $\phi_{0}$, the Fresnel propagation equation becomes,
\begin{multline}
    \Psi^\prime(x^\prime,y^\prime) = K(z=-10~\mathrm{km}) \\ \otimes \left[ \left[ K(z=+10~\mathrm{km}) \otimes \exp{(i\phi_{10})} \right] \exp{(i\phi_{0})} P(x,y) \right].
\label{eqn:Fresnel_prop}
\end{multline}
The surface layer and telescope pupil are multiplied into the wavefront before the final convolution. This is why these effects can not be de-coupled from the higher turbulent layers and the wavefront in the conjugate plane will therefore depend on the high altitude phase aberrations as well as the surface layer and will be different for the target and comparison stars. In addition to the diffraction these residual intensity variations will limit the effectiveness of the technique.

\begin{figure}
	\centering
	 \includegraphics[width=80mm]{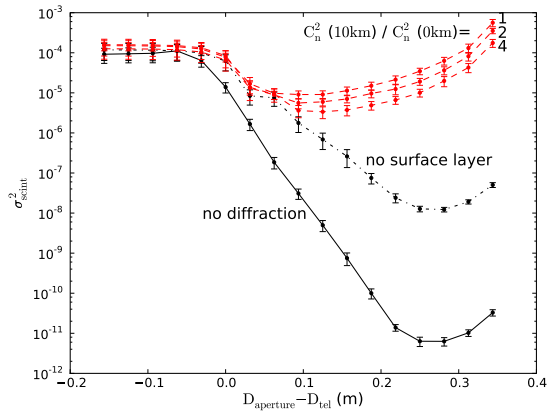}
	 \caption{The solid line shows the scintillation variance as a function of aperture size for an aperture suspended in the sky 10~km above a 2~m telescope. In this case it is possible to reduce the scintillation variance to effectively zero. The black dot--dashed line shows the scintillation variance for a single high-altitude turbulent layer with the aperture in the conjugate plane. The performance is not as good as the solid line due to the diffraction from the telescope pupil. The red dashed lines show the scintillation variance for the aperture in the conjugate plane of the high turbulent layer and with a surface layer with strengths equal to ${C}_{n}^{2} (z_0)dh$, $2\times C_{n}^{2} (z_0) dh$ and $4\times {C}_{n}^{2} (z_0) dh$, where $z_0$ is the conjugate altitude, with ${C}_{n}^{2} (z_0)dh=3.5\times10^{-13} \mathrm{m}^{1/3}$. The data points and error bars are the mean and standard errors of 20 simulations, each with unique and randomly generated phase screens.}
	 \label{fig:ap_size}
\end{figure}

Our conjugate-plane photometry concept has been simulated using a Fresnel propagation wave optics simulation using the theory stated above and randomly generated phase screens. Scintillation is often quantified by the scintillation index, $\sigma_{\mathrm{scint}}^{2}$, which is defined as the normalised variance of intensity fluctuations, $\sigma_{\mathrm{scint}}^{2} = \langle(I-\langle I\rangle)^2\rangle/\langle I \rangle^2$, where $I$ is the intensity of the image and $\langle I\rangle$ denotes the time averaged intensity \citep{Dravins1997}. Figure~\ref{fig:ap_size} shows the scintillation index as a function of aperture size for a few example cases. The first case shows the theoretical maximum reduction found by suspending the aperture in the sky above the telescope (solid line). This is entirely unfeasible but places a maximum limit on the reduction of the variance. The black dot--dashed line shows the scintillation variance for differential photometry with the aperture in the conjugate plane. Diffraction through the pupil means that light is redistributed in the pupil. Therefore, simply blocking the outer regions of the pupil will no longer remove most of the extra light and will result in a higher scintillation variance. The small shoulder in the curve at approximately 0.07~m coincides with the radius of the first diffraction ring. The red dashed lines show the scintillation variance with a high altitude layer and a surface layer which varies in strength. In this case a comparison star is required to normalise the scintillation. The strength of the surface layer is selected so that the ratio of $C_n^{2}(10~\mathrm{km})dh/C_n^{2}(0~\mathrm{m})dh$ is equal to 1, 2 and 4. If the surface layer is weaker than the high turbulent layer the residual intensity fluctuations will be lower and so the residual scintillation will also be lower.\comment{ For large apodization the scintillation variance gets worse this is because the scintillation due to the surface layer will be greater as the scintillation index is proportional to the diameter to the power $-\frac{7}{3}$ (equation~\ref{eq:scintindex1}).} The maximum median variance reduction factor for  ${C}_{n}^{2} (10~\mathrm{km})dh/{C}_{n}^{2} (0~\mathrm{km})dh = 1$ (i.e. equal strength), 2 and 4 is 17, 23 and 47, respectively and is found at $D_{\mathrm{aperture}} - D_{\mathrm{tel}}\approx0.1~\mathrm{m}$, for a simulated telescope diameter of 2~m.

The amplitude of the first diffraction ring is substantially larger than any others (as seen in figure~\ref{array:scint_pupil}). The optimum aperture size is therefore one which blocks this ring but none of the others. This will minimise the residual diffraction and retain a large pupil area. \comment{A smaller aperture would block more of the diffraction rings more however as $\sigma_{\mathrm{scint}}^{2} \propto D_{\rm{tel}}^{-7/3}$ it would also increase the residual scintillation from all other turbulent layers, including the surface layer.} The radius of the first diffraction ring in the very near field is given by the radius of the first Fresnel zone, $r_\mathrm{F}=\sqrt{\lambda z}$, in this case 0.07~m and is independent of telescope size.

\begin{figure}
	\centering
	 \includegraphics[width=80mm]{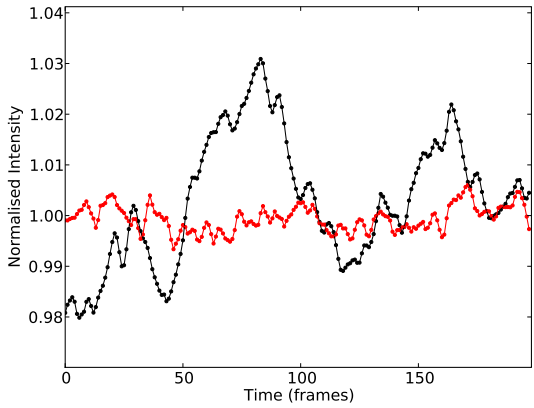}
	 \caption{An example simulated light curve. The black line is the intensity pattern from a simulation observing a star with a 2~m class telescope through the atmosphere with a turbulent layer at 10~km and 0~m, both with $\int C_n^2 dh = 353\times10^{-15}~ \rm{m}^{1/3}$. The exposure time of each frame is short so that there is no temporal averaging of scintillation. The red line shows the scintillation corrected light curve. In this case the intensity variance is reduced from $1.6\times 10^{-4}$ to $6.4 \times 10^{-6}$, a factor of 25. The residual noise is due to the uncorrected scintillation.}
	 \label{fig:calibrated_intensity}
\end{figure}

The reduction in scintillation noise can be clearly seen in figure~\ref{fig:calibrated_intensity}, which shows the normalised light curve for a sequence of 200 frames from a simulation assuming a constant source intensity. The black line shows the original light curve with a variance of $1.6\times 10^{-4}$ due to scintillation. The red line is the light curve after scintillation reduction and has a variance of $6.4 \times 10^{-6}$, a reduction factor of 25. The variance is in units of normalised intensity, $\Delta I/ I$. The simulation assumes an atmosphere with two turbulent layers, one at the ground and one at 10~km, both with $\int C_n^2 dh = 353\times10^{-15}~ \rm{m}^{1/3}$, the telescope diameter was 2~m and there was no temporal averaging.

\begin{figure}
	 \includegraphics[width=80mm]{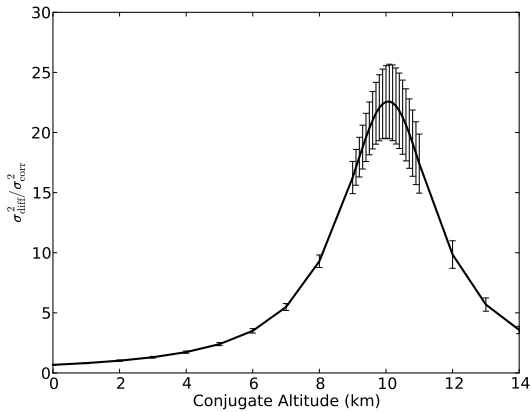} \\
\caption{Ratio of intensity variance for normal differential photometry ($\sigma^2_{\mathrm{diff}}$) and scintillation corrected photometry ($\sigma^2_{\mathrm{corr}}$) versus conjugate altitude for an atmosphere with layers at 0~m and 10~km, both with $\int C_n^2 dh = 353\times10^{-15}~ \rm{m}^{1/3}$ and a telescope diameter of 2 m. The curve is Lorentzian with a FWHM of approximately 3.5~km. At conjugate altitude 0~m we measure an improvement in the intensity variance of $\sim$0.5, i.e. the variance is actually increased. This is because the pupil size is reduced by the apodizing mask. The data points and error bars are the mean and standard deviation of 20 simulations, each with unique and randomly generated phase screens.}
	 \label{fig:conj_alt}
\end{figure}

A mis-conjugation of the aperture will result in less than optimal performance. Figure~\ref{fig:conj_alt} shows the factor by which the scintillation variance is reduced as a function of conjugate altitude for turbulent layers at 0~m and 10~km. \comment{The curve is Lorentzian and}In this case the curve has a full width at half maximum of approximately 3.5~km. This will be narrower for turbulent layers at lower altitudes and wider for higher altitudes. Knowledge of the contemporaneous turbulence profile is therefore essential to ensure that the aperture is conjugate to the correct altitude. 

\section{Performance Estimation}

The Monte-Carlo simulations are useful to examine the performance for a particular parameter set. However, it is very inefficient for modelling the performance as a function of time for real turbulence profiles with many turbulent layers. To do this an analytical estimate of the intensity variance for a given parameter set is required. 

If the pupil is much larger than the Fresnel radius ($D \gg \sqrt{\lambda z_{0}}$) the intensity variance due to scintillation, $\sigma^{2}_{\mathrm{scint}}$, can be predicted using the theoretical model described by \citet{Dravins1997b},
\begin{equation}
\sigma^{2}_{\mathrm{scint}} \propto {D_{\rm{tel}}^{-\frac{7}{3}} (\sec{\gamma})^{3} \int_{0}^{\infty} C_{n}^{2}(h) h^{2} dh  },
\label{eqn:scint_var}
\end{equation}
where $\gamma$ is the zenith angle. The scintillation index is then independent of wavelength and proportional to the altitude of the turbulent layer squared and the strength of the turbulent layer. We can calculate the scintillation index due to all of the turbulent layers assuming the pupil is conjugate to an altitude, $z_{0}$. In this case the scintillation index, $\sigma_{\mathrm{z_{0}}}^{2}$, at a given altitude can be calculated using a modification to the scintillation index equation (equation~\ref{eqn:scint_var}),
\begin{equation}
	\sigma_{\mathrm{z_{0}}}^{2} \propto D_{\rm{tel}}^{-\frac{7}{3}} (\sec{\gamma})^{3} \int_{\mathrm{SL}}^{\infty} {C_{n}^{2} \left(h\right) \left(h-z_{0}\right)^{2} dh},
\end{equation}
where $(h-z_{0})$ is the separation between the layer altitude and the conjugate altitude, ignoring the surface layer as this will be dealt with separately.

The corrected residual scintillation variance, $\sigma^{2}_{\mathrm{corr}}$, will be dominated by this but we also add noise terms due to the pupil diffraction and the surface layer. These noise sources are independent but the total is modulated by the original scintillation variance (equation~\ref{eqn:Fresnel_prop}) and so the total residual scintillation variance can be modelled by,
\begin{equation}
\sigma_{\mathrm{corr}}^{2} = 2\sigma_{z_{0}}^{2} + \left( (\sigma_{\mathrm{scint}}^{2}) ^{j} \times \left((\sigma_{\mathrm{SL}}^{2}) ^{k} + F^{l}\right) \right),
\label{eqn:corrected_scint_var}
\end{equation}
\comment{overlap problem!!}
where $\sigma_{\mathrm{SL}}^{2}$ is the scintillation index due to the surface layer, $F$ is the Fresnel number used to quantify the `amount' of diffraction and is given by $F=D^2/4\lambda z$, and $j$, $k$ and $l$ are solved empirically from the simulation results and are found to be $j = k = 2/3$, $l=-1.4$.

\begin{figure}	
	 \includegraphics[width=80mm]{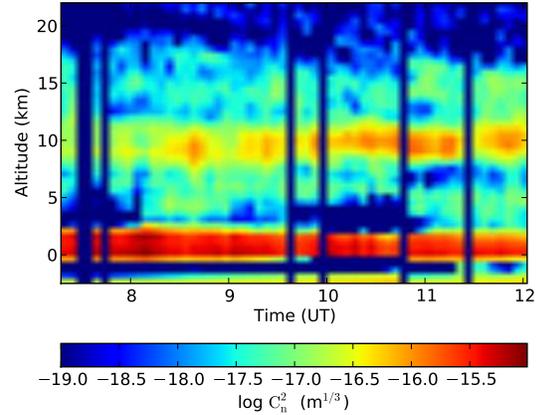} \\
 	\caption{SCIDAR turbulence profile, i.e. height above sea level against time, where the colour indicates the strength of the turbulence, from 2000 May 19 at San Pedro M$\acute{\mathrm{a}}$rtir. The profile shows a dominant layer at approximately 10~km throughout the night. San Pedro M$\acute{\mathrm{a}}$rtir is located at 2800~m above sea level.}
	 \label{fig:scidar_prof_2000_05_19}
\end{figure}

\begin{figure}	
	  \includegraphics[width=80mm]{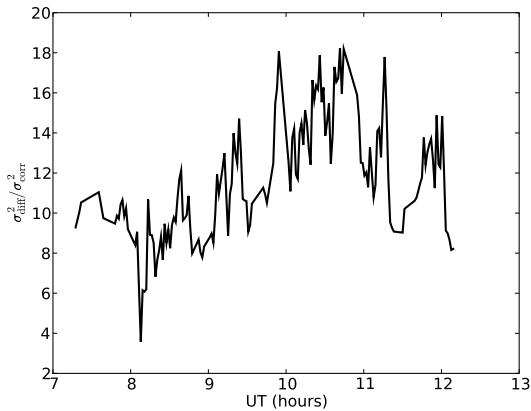} 
 	\caption{Improvement in intensity variance as a function of time for the night of 2000 May 19. The median improvement ratio for this night is $\sim$11.5.}
	 \label{fig:scidar_improvement}
\end{figure}

Using high-resolution generalized SCIDAR turbulence profile data from San Pedro M$\acute{\mathrm{a}}$rtir \citep{Avila06} and the model developed from the simulation results we can estimate the expected improvement in intensity variance. The SCIDAR profile shown in figure~\ref{fig:scidar_prof_2000_05_19} was recorded on 2000 May 19 and shows a strong turbulent layer at approximately 10 km throughout the night. Figure~\ref{fig:scidar_improvement} shows the expected improvement factor in intensity variance as a function of time for the same night. The median improvement ratio is 11.5 for this example.

When calculating expected performance for real experiments it is also necessary to include the exposure time of the integration as this will average out the scintillation and reduce the intensity variance. The scintillation index given in equation~\ref{eqn:scint_var} is only valid for very short exposures where there is no temporal averaging, i.e. the exposure time has to be less than the pupil crossing time of the intensity fluctuations. The crossing time, $t_c$, can be calculated as $t_c = D_{\rm{tel}}/v_w$, where $v_w$ is the velocity of the turbulent layer. If the exposure time, $t$, is greater than the crossing time then the scintillation index is modified to \citep{Kenyon06},
\begin{equation}
	\sigma_{\mathrm{scint}}^{2} \propto \frac{D_{\rm{tel}}^{-4/3}}{t} \int{ \frac{ C_{n}^{2} \left(h\right) h^{2} } {V\left(h\right)}  dh},
\end{equation}
where $V(h)$ is the velocity of the turbulent layer at altitude $h$. Using this modification to the scintillation index we can calculate an example light curve for a fictional extrasolar planet transit for a given turbulence profile.

\begin{figure}
	\centering
	 \includegraphics[width=80mm]{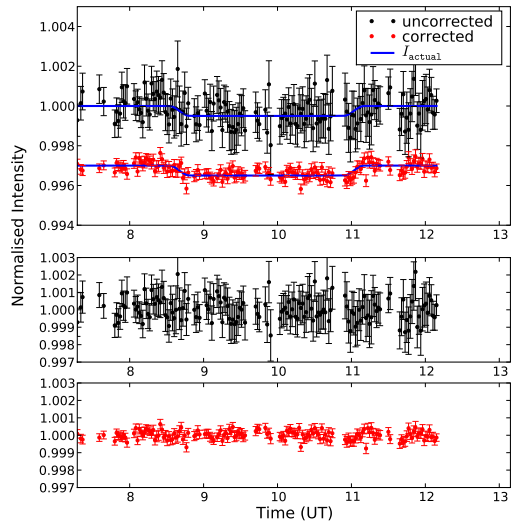}
	 \caption{Simulated light curve of the secondary transit of an extrasolar planet with a 0.05\% transit depth. The data were calculated assuming the same atmospheric parameters as measure by SCIDAR (figure~\ref{fig:scidar_prof_2000_05_19}) and a 2.0 m telescope with 30~s exposure times in the v-band and a target magnitude of 11. The top panel show the simulated light curves with no scintillation correction (black points, top) and with scintillation correction (red points, bottom), offset for clarity. The solid lines show the theoretical light curve (i.e. with no noise). The data points are randomly selected from a distribution with a variance equal to the total noise at that time, and the error bar indicates the total noise at that time. The lower panels show the normalised residuals.}	 
	 \label{fig:planet_example}
\end{figure}

Figure~\ref{fig:planet_example} shows an example simulated extrasolar planet transit. The transit depth is assumed to be 0.05~\% and has a duration of 2.5 hours. A 2~m telescope and 30~s exposure time are also assumed. The optical turbulence profile used in the simulation is the same as that shown in figure~\ref{fig:scidar_prof_2000_05_19} as measured by SCIDAR at San Pedro M$\acute{\mathrm{a}}$rtir. A wind speed of 5~ms$^{-1}$ for the surface layer and 20~ms$^{-1}$ for all other turbulence is assumed. The normalised scintillation noise in the visible is reduced from $0.70\times 10^{-3}$ (0.78~mmag) to $0.21\times 10^{-3}$ (0.23~mmag), an improvement factor of 3.3. If we assume a target magnitude of 11 then we have reduced the scintillation to a level which is comparable to the shot noise.\comment{ The total noise limits the detectable transit depth. The reduction in scintillation noise shown in figure~\ref{fig:planet_example} will allow 3$\sigma$ detections of transit depths from 0.3\% to 0.06\%.}

\begin{figure}
	\centering
	 \includegraphics[width=80mm]{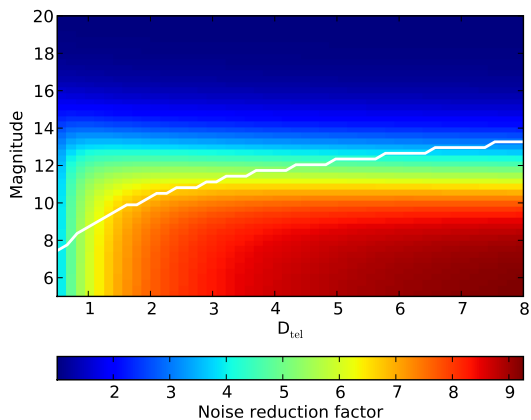}
	 \caption{The magnitude of the improvement we can expect to observe with conjugate-plane photometry depends on the other noise sources. If we assume the same parameters as in figure~\ref{fig:planet_example} and an atmosphere given by the median SCIDAR profile then we can plot the noise reduction factor as a function of target magnitude and telescope diameter. The white line indicates the limiting magnitude for a given telescope size to prevent saturating a 16-bit analogue to digital converter in a CCD. The optimum telescope size is therefore the maximum noise reduction factor just above this curve, i.e. between 1.2~m and 2~m. This will vary with seeing and camera parameters.}	 
	 \label{fig:mag_D_noise}
\end{figure}

Although the aperture must be placed at the conjugate altitude of the turbulence the photometry can be done in the focal plane. This means that we do not expect any of the other noise sources to increase as a result of implementing our conjugate-plane photometry technique. The magnitudes of other noise sources, such as shot noise, readout noise or flat fielding noise, will depend on other factors. There are three possible regimes in which we are interested: scintillation dominated, other noise dominated and a mixture of the two. In the first and last cases the noise will add in quadrature and so a reduction in scintillation noise by a factor of $n$ will reduce the total noise to, $\sigma_{T_{2}}=\sqrt{\sigma_{T}^{2} + \sigma_{\mathrm{scint}}^{2}\left(\frac{1}{n^2}-1\right)}$, where $\sigma_{T}$ is the total noise before scintillation reduction. Figure~\ref{fig:mag_D_noise} shows a 2D plot of the total noise reduction factor as a function of the telescope diameter and the target magnitude assuming the same parameters as before. The atmospheric model was the median profile from the SCIDAR data recorded on 2000 May 19. The optimum telescope size is found to be between 1.2~m and 2~m. Less than this and the diffraction effects limit the possible scintillation noise reduction and apertures greater than this become shot noise dominated. In the latter scenario the scintillation noise is insignificant and so scintillation correction techniques will have no effect.

The median reduction in intensity variance for all available SCIDAR data collected over 24 nights in March/April 1997 and May 2000 at San Pedro M$\acute{\mathrm{a}}$rtir is a factor of 6. However, with the limited data available it is difficult to say if this representative; it is possible that other times or sites will yield even better results if the turbulence is more constrained to stratified layers.

Adaptive optics (AO) can be used to reduce the phase aberrations for imaging. Here it is intensity fluctuations which are the problem and so AO systems can not directly reduce the scintillation. However, AO systems can be used in conjunction with this technique to further reduce the scintillation. As shown previously the surface turbulent layer is a major limitation to the conjugate-plane technique. Therefore, a ground layer adaptive optics (GLAO) system could be used to remove the phase aberrations induced by the turbulent surface layer and therefore also reduce the residual scintillation. On occasions when the atmosphere is dominated by a number of turbulent layers a multi-conjugate AO system \citep{Langlois04} combined with conjugate plane masks could be used to significantly reduce the scintillation.

\section{Opto-mechanical design}
\begin{figure}
	\centering
	 \includegraphics[width=70mm]{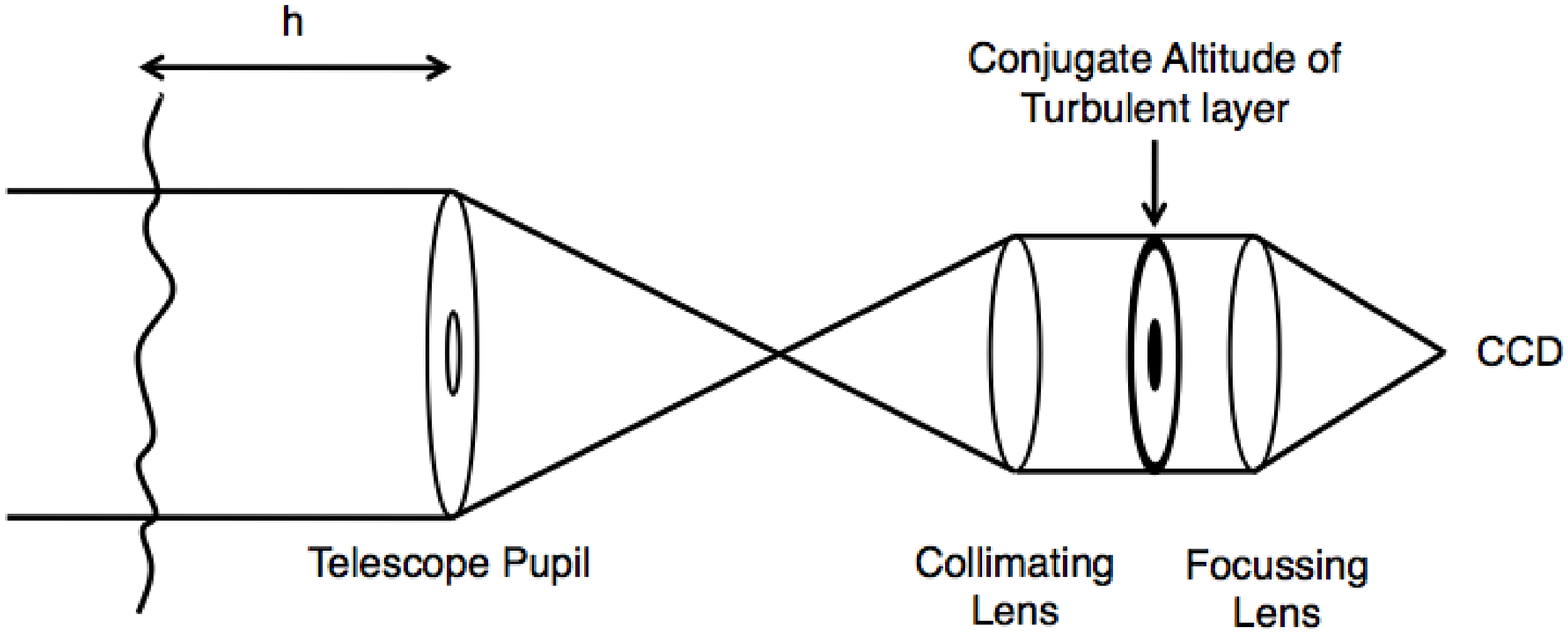}
	 \caption{Conceptual design for one arm of the instrument.}\comment{The physical design of the instrument is actually very simple. An aperture is placed is placed at the conjugate altitude of the turbulent layer and the light is refocused onto the CCD in the focal plane. However, as the aperture is not in the pupil plane off-axis stars will not all illuminate the same area of the aperture. This means that separate arms are required for the target and comparison stars.}
	 \label{fig:apod_opt}
\end{figure}

\begin{figure}
	\centering
	 \includegraphics[width=80mm]{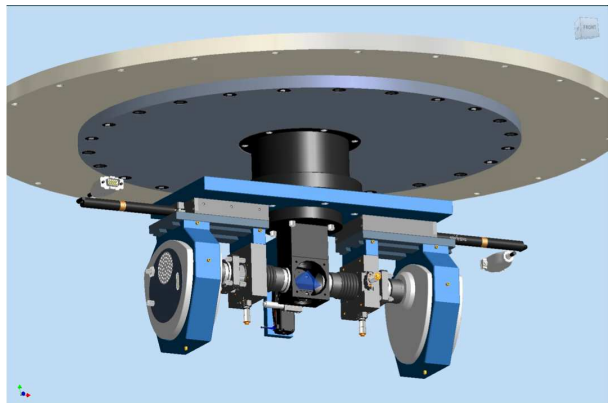}
	 \caption{Prototype of the conjugate-plane photometer that we are due to test on-sky shortly.}	 
	 \label{fig:mech_design}
\end{figure}

The design of a conjugate-plane photometer is actually very simple. Figure~\ref{fig:apod_opt} is a diagram of such an instrument instrument. An aperture is placed in the collimated beam at the conjugate plane of the turbulent layer. A lens is then used to focus the light onto a CCD in the focal plane. As the aperture is not in the pupil plane, any off-axis light will not illuminate the whole aperture and therefore a separate optical arm is required for the target and comparison stars. This can be achieved with either a prism near the focal point of the telescope, or with pick off mirrors if more stars are required. This is completely different to an adaptive optics type approach as there are no moving parts once the altitude has been set. 
Figure~\ref{fig:mech_design} shows the full design of a prototype instrument, which we shall shortly be commissioning to demonstrate the conjugate-plane photometry technique.

\section{Conclusions}
We have presented a technique, known as conjugate-plane photometry, to improve the precision of fast photometry from ground based telescopes. The dominant source of noise from the Earth's surface is often scintillation due to high altitude turbulent layers. By placing an aperture at the conjugate altitude of this layer we can remove the majority of the scintillation from this layer. We still detect scintillation from other layers, but evidence from turbulence profile measurements suggests that at premier observing sites the atmosphere typically consists of a single strong high-altitude layer and a strong boundary layer. Under such condition our technique could remove a large fraction of the scintillation. Simulations show that the intensity variance can be reduced by an order of magnitude. Theoretical calculations have been used to estimate the scintillation noise reduction for a given parameter set. For example, with an atmosphere as measured by SCIDAR at San Pedro M$\acute{\mathrm{a}}$rtir on the 19$^\mathrm{th}$ May 2000, the median reduction in intensity variance is a factor 11.5 . Using all available SCIDAR data including times when we do not see a dominant high altitude layer we still obtain a median improvement of a factor of 6. This is because we are reducing the propagation distances from any single layer to the conjugate altitude and the scintillation index is proportional to propagation distance squared. By generating a synthetic light curve for a 2~m telescope in the visible using the variance expected from SCIDAR data and exposure times of 30~s it was found that we could reduce the scintillation noise from 0.78~mmag to 0.21~mmag, comparable to the shot noise. This reduction in noise will open up new science areas from the ground, including the characterisation of extrasolar planets through the observations of the secondary transit.\comment{, resulting in a 3$\sigma$ detection for systems with a 0.06~\% dip in intensity. } The conjugate-plane photometer is easy to implement as a passive correction technique. However, it does require a contemporaneous SCIDAR measurement in order to ensure the aperture is at the correct plane.

\section*{Acknowledgements}
We are grateful to the Science and Technology Facilities Committee (STFC) for financial support (JO). RA acknowledges financial support from CONACyT and PAPIIT through grants number 58291 and IN107109-2.
\bibliographystyle{mn2e}
{}

\end{document}